\documentclass[ aps,%
floatfix,%
final,%
notitlepage,%
oneside,%
onecolumn,%
nobibnotes,%
nofootinbib,%
superscriptaddress,%
showpacs,%
showkeys
]%
{revtex4}

\usepackage{graphicx}

\begin{document}

\title{A search for  decay $\eta' \rightarrow 4 \pi^{0}$ with GAMS-$4\pi$ Setup}

\author{S.V.~Donskov}
\author{V.N.~Kolosov}
\author{A.A.~Lednev}
\author{Yu.V.~Mikhailov}
\author{V.A.~Polyakov}
\author{V.D.~Samoylenko}
\email{Vladimir.Samoylenko@ihep.ru}
\author{G.V.~Khaustov}
\affiliation{Institute for High Energy Physics, Protvino, Russia}

\begin{abstract} 
 A  search for rare decay $\eta' \rightarrow 4 \pi^{0}$  has been performed
with  GAMS-4$\pi$ Setup. The new upper limit for decay was
established $BR(\eta' \rightarrow 4 \pi^{0}) < 3.2 \cdot 10^{-4}$ at 90\% confidence level.
The $\pi^{-} p$ charge-exchange reaction at 32.5 GeV/c was used 
as a source of $1.3\cdot 10^{6}$ $ \eta'$ mesons. 
Experiment carried out at the IHEP U-70 accelerator.
\end{abstract}
\pacs{
13.25.Jx,	%Decays of other mesons
13.75.-n,
13.75.Lb
}
\keywords{CP violation, rare decay, eta', pseudoscalar}
\maketitle 
\section{Introduction}

\begin{figure}[htb]
%\begin{center}
%\begin{picture}(120,45)
%\centerline{\epsfig{file=fig2g_pik.eps,height=12.5cm,width=15cm}}
%\put(-10,2){
%\includegraphics[width=0.90\textwidth]{diags2-1.eps}
\includegraphics[width=0.90\textwidth]{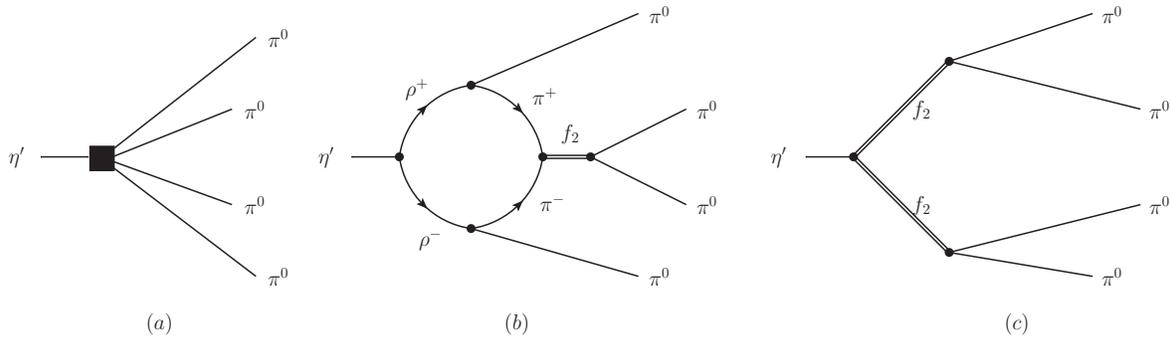}
%}
%\put(10,2){{ \it a}}
%\put(50,2){{\it b}}
%\put(100,2){{\it c}}
%\end{picture}
%\end{center}
\caption{Decay $\eta' \rightarrow 4 \pi^{0} $ with direct $CP$ violation (a) ;
with $\pi^{+} \pi^{-} \rightarrow \pi^{0} \pi^{0}$
rescattering, including intermediated $f_{2}$ meson (b);
by virtual state $f_{2} f_{2}$  in $P$ wave (c).
}
\label{fig_diag}
\end{figure}

Assuming $CP$ conservation in strong interaction the direct decay $\eta' \rightarrow 4 \pi^{0}$
is forbidden, Fig.~\ref{fig_diag}a. 
From other point of view  decays with odd number of pseudoscalars  
belong to the wider class of anomalous decays,
in low-energy QCD such anomalous decays are governed by the Wess - Zumino - Witten (WZW) term \cite{wess}.
In this approach the decay probability is not vanish, and  there are as $CP$-conserving and $CP$- 
violating models for the process $\eta' \rightarrow 4 \pi^{0} $ \cite{parashar, wirzba, guo}.
%
%and the different variants for these decays are discussed in \cite{parashar, wirzba, guo}.
%

For $CP$-conserving case the decay amplitude is enhanced by pion loop contribution 
shown in Fig.~\ref{fig_diag}b 
and by $f_{2}f_{2}$ virtual intermediated state, shown in Fig.~\ref{fig_diag}c.
The estimate of a decay branching   in this approach
leads to  $BR( \eta' \rightarrow 4 \pi^{0}) \approx 4 \cdot 10^{-8}$.

One would  hope that $CP$-violating variant is more preferable
to avoid  huge angular-momentum suppression. $CP$-violating mechanism is introduced
by $\theta$-term  in QCD Lagrangian to solve $U(1)_{A}$ problem. Carefull consideration of
this method in Ref.\cite{pich} shows that the branching ratio is 
$BR (\eta' \rightarrow 4\pi^{0}) \approx 0.1 \cdot \theta^{2} $.
Current limit for $\theta$-term based on neutron electric dipole moment (EDM) measurements
is $\theta < 10^{-11} $, and this value makes the branching far from modern experimental
possibilities. The deep connection between neutron EDM  and $CP$-violating decays of $\eta'$ is 
discussed in Ref.\cite{gorshtein}.

Here we present the result of a new search of the decay
\begin{equation}
\eta' \rightarrow 4 \pi^{0}.
\label{etap4pi}
\end{equation}
Previous result was published by GAMS Collaboration with upper limit
$BR(\eta' \rightarrow 4 \pi^{0}) < 5 \cdot 10^{-4}$
at 90\% confidence level \cite{gams_etap, pdg}.
   
% Eur.Phys.J. C73 (2013) 2614
\section{GAMS-$4\pi$ Setup and Event selection}

The experiment was carried out  at the U70 IHEP accelator, in
a secondary beam  of negative particles.
The type of a beam particle is defined reliably by two threshold Cherenkov 
counters \cite{pps}  with quartz optics. 
The interaction point of the beam particles is measured with precision
$\pm 4$ cm from the amount of Cherenkov light, 
emitted in  the 40 cm long liquid hydrogen target.
The target is surrounded by guard system, which  consists of scintilation veto layer
and lead glass counters  to detect recoiled baryons.
This set of  the detectors allows the  identification of the charge-exchange processes
with neutral final states.
The main detector of GAMS-$4\pi$ Setup is the lead glass electromagnetic calorimeter
GAMS \cite{gams}.
The central part of the calorimeter contains  PWO crystals in order 
to provide good energy and spatial resolution \cite{pwo}. 
A more detailed description of the performance of the experiment and data processing 
has been given elsewhere  \cite{gams_epj, gams4pi}.
 
The charge-exchange reaction
\begin{equation}
%\pi^{-} \mbox{ } p \rightarrow \eta' \mbox{ }n 
\pi^{-} \mbox{ } p \rightarrow M^{0} \mbox{ }n 
\label{cex}
\end{equation}
with a 32.5 GeV/c   beam momentum was
used as the source of the monoenergetic $\eta'$ mesons

\begin{figure}[htb]
%\begin{center}
%\begin{picture}(140,80)
%\centerline{\epsfig{file=fig2g_pik.eps,height=12.5cm,width=15cm}}
%\put(2,2){
\includegraphics[width=0.90\textwidth]{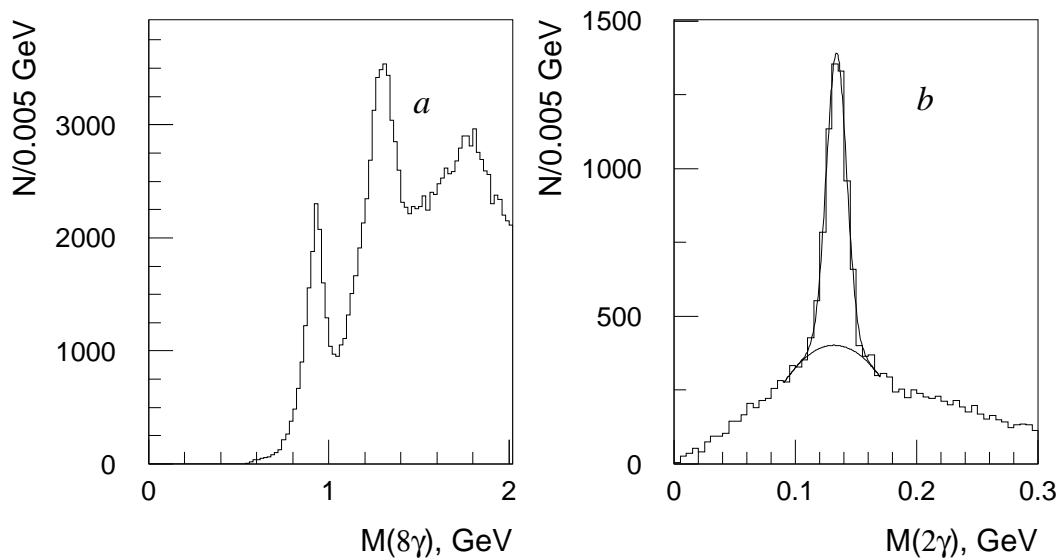}
%}
%\put(60,60){{ \it a}}
%\put(125,60){{\it b}}
%\put(60,50){{\it в}}
%\%put(135,50){{\it г}}
%\end{picture}
%\end{center}
\caption{Experimental mass spectrum of $8\gamma$ events after 1C-fit ({\it a});
Mass spectrum of fourth non-fitted $\gamma$-pair after 4C-fit for events
with $M_{8\gamma} < 1.2$ GeV ({\it b}). Spectrum was fitted by gaussian and 2nd order
polynomial, mass $m_{\pi^{0}} = 0.134$ GeV, width $\sigma_{\pi^{0}} = 0.008$ GeV.
}
\label{fig_8g_pi}
\end{figure}

The analysis of  the data was performed in several steps.
At first the 4-momenta of the photons were  reconstructed 
by special procedures  \cite{lednev1,lednev2} 
and events with exactly eigth photons in GAMS were selected.
Next, a series of criteria were applied in order to suppress
instrumental and physical backgrounds without losing
detection efficiency of the reactions under study (\ref{cex}):
%
%\begin{enumerate}
\begin{itemize}
\item The distance between the shower centres in GAMS is larger
than 40 mm;
\item The distance between photon's impacts in GAMS
and the beam axis is larger than 40 mm. This requirement
provides suppression of the background in the central part of calorimeter
where the beam is most intense;
\item The total energy deposited in GAMS is limited to the range
(28 -- 36) GeV;
\item The energy of each photon is larger than 0.5 GeV.
%\end{enumerate}
\end{itemize}
Mass spectrum of $8\gamma$ events after 1C-fit (fixing the mass of the  recoiled 
neutron in reaction~(\ref{cex}), fit probability CL$ > 0.05$)  is shown in Fig.~\ref{fig_8g_pi}a.

$4\pi^{0}$ system is rather difficult for experimental study, and was studied before
in the reports  \cite{na12,kek} dealing with it in charge-exchange reaction. 
%
%The physical background originates from limited spatial and energetic resolution
%of the calorimeter, which is enhanced by combinatorics for eight $\gamma$.
%
To estimate the background in low mass region 4C-fit was performed (fixing neutron
and three $\pi^{0}$ masses). The non-fit $\gamma$-pair mass is shown in Fig.~\ref{fig_8g_pi}b,
and one can see that in this hard kinematic region the integral ratio signal/background
for mass interval $(0.11-0.16)$ GeV  is $S/B = 1.1$.
\begin{figure}[htb]
%\begin{center}
%\begin{picture}(140,90)
%\centerline{\epsfig{file=fig2g_pik.eps,height=12.5cm,width=15cm}}
%\put(2,2){
\includegraphics[width=0.90\textwidth]{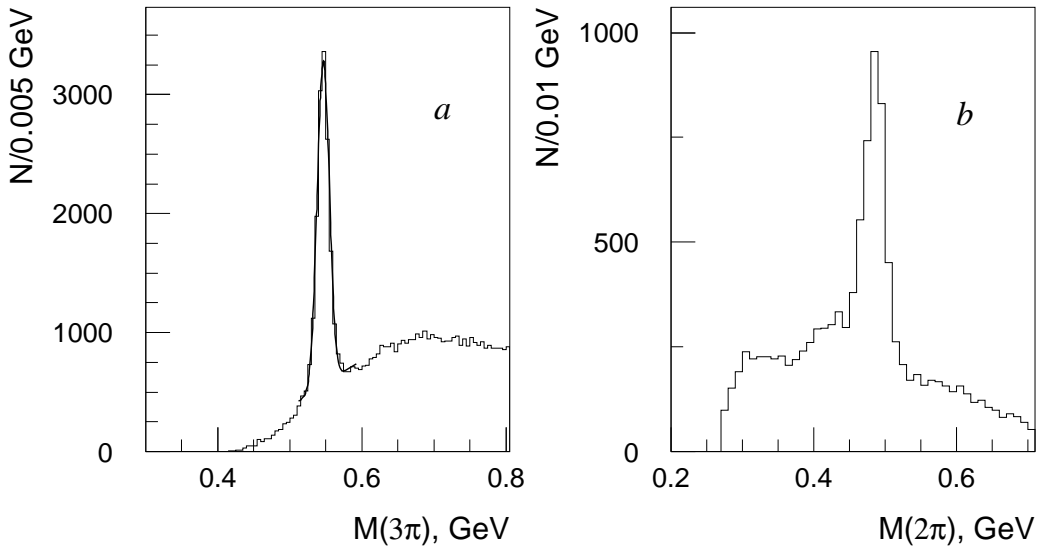}
%}
%\put(60,60){{\it а}}
%\put(125,60){{\it б}}
%\put(60,50){{\it в}}
%\%put(135,50){{\it г}}
%\end{picture}
%\end{center}
\caption{Mass spectrum  $3\pi^{0}$ subsystem, 4 ent./event ({\it a}).
Fitted mass of $\eta$  meson is $m_{\eta}=0.547$ GeV and width $\sigma_{\eta} = 0.008$ GeV;
Mass spectrum  $2\pi^{0}$ subsystem, when the mass other $\pi^{0} \pi^{0}$ pair
belongs to  (0.46 -- 0.52) GeV region ({\it b}).
}
\label{figetaks}
\end{figure}

$\eta \pi^{0}$ system produced with high intensity, and in case 
of the decay mode  $\eta \rightarrow 3\pi^{0}$
has four $\pi^{0}$ mesons  in final state. Such $\eta$-mesons are seen clearly, Fig.~\ref{figetaks}a. 
To suppress this physical background 5C-fit (fixing neutron and four $\pi^{0}$ masses) 
was performed, and if the mass of any $3\pi^{0}$ subsystem belongs to
$\eta$ meson interval  the event is rejected.
Pair $\pi^{0}$ mass spectrum consists of $K^{0}_{s}$ mesons,  Fig.~\ref{figetaks}b, but the contribution 
$K_{s}^{0} K_{s}^{0}$ system at mass $\approx 1$ GeV is negligible. 

\begin{figure}[htb]
%\begin{center}
%\begin{picture}(100,85)
%\put(2,2){
\includegraphics[width=0.9\textwidth]{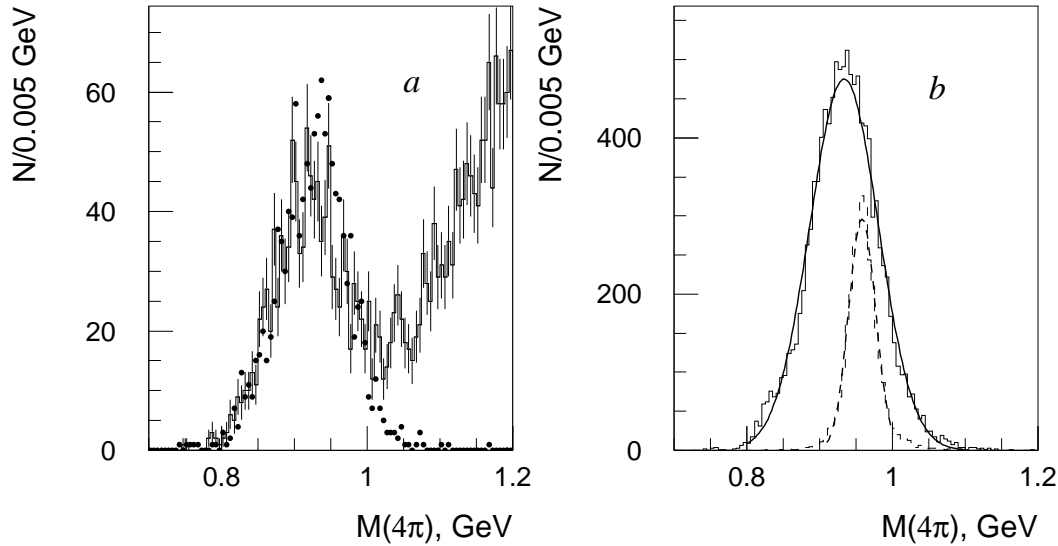}
%}
%\put(60,70){{\it а}}
%\put(135,70){{\it б}}
%\end{picture}
%\end{center}
\caption{Final experimental $4\pi^{0}$ mass spectrum. Points are Monte-Carlo events 
for decay chain (\ref{etap5pi}) (a);
Monte-Carlo spectra for decay  (\ref{etap5pi}) (solid histogramm) and for  (\ref{etap4pi}) (dashed histogramm),
as an illustration (b).
}
\label{fig_4pi}
\end{figure}

Pure $4\pi^{0}$ event mass spectrum is presented  in Fig.~\ref{fig_4pi}a and exhibit
a wide peak with mass around 0.9 GeV. The origin of this structure can be explained
by decays chain
\begin{equation}
\eta' \rightarrow \eta \pi^{0} \pi^{0}, \mbox{ }
\eta \rightarrow 3 \pi^{0}
\label{etap5pi}
\end{equation}
with two lost $\gamma$. 
Taking into account that 105 possible combinations exist to compose $4\pi^{0}$ from $8\gamma$, 
%and limited spatial and energetic resolution of the calorimeter 
the remaining eight photons  can easy form $4\pi^{0}$ system.

The method proposed in Ref.~\cite{barlow} was used to set upper limit for the decay desired.
The decays (\ref{etap4pi}) and (\ref{etap5pi}) are consider as the event sources 
for the experimental histogramm, Fig.~(\ref{fig_4pi}). The mass spectra for these decays 
have been obtained by Monte-Carlo simulation.  The mass shift of $\eta'$  peak position
for decay (\ref{etap5pi}) with two lost $\gamma$ is $\approx 30$~MeV, 
the peak becomes the width $\sigma_{5\pi} \approx 50$~MeV, Fig.~\ref{fig_4pi}b.
For decay (\ref{etap4pi}) the mass resolution is $\sigma_{4\pi} = 20$~MeV  after 5C-fit.
One can see in Fig.~\ref{fig_4pi}a that experimental $4\pi^{0}$ mass spectrum in (0.8--1.0) GeV interval 
is reproduced well by background spectrum (\ref{etap5pi}).
The upper limit is evaluated as  
\begin{equation}
BR(\eta'  \rightarrow 4\pi^{0}) <  \frac{N_{exp} \dot  (W (\eta' \rightarrow 4\pi^{0}) + dW(\eta' \rightarrow 4\pi^{0}))}
{ \epsilon  (\eta' \rightarrow 4\pi^{0}) \dot N_{\eta'} },
\label{uplimit}
\end{equation}
where $W (\eta' \rightarrow 4\pi^{0})$ is the weight of desired process (\ref{etap4pi}),
$dW(\eta' \rightarrow 4\pi^{0}) $ is the error corresponding 90\% confidence level
in generalized likelihood method\footnote{subroutine HMCMLL in HBOOK},
$N_{exp}$ is number of the events in the experimental histogramm region under investigation,
$N_{\eta'}$ is number of $\eta'$ mesons and $\epsilon$ is detection efficiency for the decay
(\ref{etap4pi}).  No signal for the decay is observed. 
At 90\% confidence level the upper limit is
\begin{equation}
BR(\eta' \rightarrow 4\pi^{0}) <3.2 \cdot 10^{-4}.
\label{reslimit}
\end{equation}
%
%-----------------------------------------------------------
%
\section{Conclusion}

We have presented our result for  a search the decay $\eta' \rightarrow 4\pi^{0}$.
One was found that main physical background arises from the chain of other neutral $\eta'$ decays
with lost photons. This process can present a problem for other experiments, such as
Crystal Ball \cite{cball}, Crystal Barrel \cite{cbarr} и WASA \cite{wasa} to find
the decay (\ref{etap4pi}).
To reduce this background electromagnetic calorimeter used in experiment
must have a low detection threshold for photons and reconstruction procedures capable to resolve
overlapping showers.

\section{Acknowledgement}
The physical problems related to the decay $\eta' \rightarrow 4\pi^{0}$ were discussed
with A.K.~Likhoded, A.V.~Luchinsky, V.V.~Kiselev, V.F.~Obraztsov, G.P.~Pronko
and A.L.~Kataev, V.A.~Rubakov (INR, Moscow).
We are grateful to all of them for interest to our work.
\end{document}